\pdfoutput=1
\documentclass{appolb}

\usepackage[utf8]{inputenc}
\usepackage[T1]{fontenc}

\usepackage{graphicx}
\usepackage{subcaption}
\usepackage{lmodern}
\usepackage[hidelinks]{hyperref}
\begin{document}
\title{
\begin{flushright}
\scriptsize{MPP-2015-271}
\end{flushright}
Introducing Loopedia%
\thanks{Presented at the XXXIX International Conference of Theoretical Physics “Matter to
the Deepest”, Ustroń, Poland, September 13–18, 2015.}%
}
\author{Viktor Papara
\address{Max-Planck-Institute for Physics\\
F\"ohringer Ring 6, 80805 M\"unchen, Germany}
% \\
% {Third Author of different affiliation
% }
% the Name(s) of other Author(s)
% \address{affiliation}
}
\maketitle
\begin{abstract}
Loopedia is a new project for a public database of loop integrals. After stating
the goals and desired properties of the project, two possible ways to
characterize Feynman graphs are explained here: the Adjacency List and the
Nickel Index. Furthermore, a first version of the Loopedia website is presented.
It allows for automatic graph generation and the conversion from Adjacency List
to Nickel Index.
\end{abstract}
\PACS{     % https://www.aip.org/pacs
02.10.Ox,  % Combinatorics; graph theory
03.65.Ca   % Formalism
}
\section{Introduction}
Loop integrals play an important role in particle physics --- especially for
high-precision calculations. For one-loop integrals there are already public
tools and repositories such as: \texttt{LoopTools}~\cite{Hahn:1998yk},
\texttt{Golem95}~\cite{Binoth:2008uq},
\texttt{PJFry}~\cite{ValeryPhD,Fleischer:2010sq},
\texttt{QCDloop}~\cite{Ellis:2007qk}, \texttt{ONELOOP}~\cite{vanHameren:2010cp},
and \texttt{Hepforge}~\cite{Buckley:2006nm}. For two-loop and multi-loop
integrals, however, there is no one-stop repository yet.

Loopedia aims to improve this situation. It will be a public database of loop
integrals. The infrastructure is provided by the contributors of the project in
a way that allows the physics community to upload integrals. Interested readers
are invited to play with the existing Loopedia Web interface now and contribute
integrals once it is fully online.

Loopedia stores the results for the integral along with the references to the
work where it was computed. This will help researchers to find results from the
literature.
\section{Characterization of graphs}
Loop integrals are closely related to Feynman graphs. This allows us to use
graph-theoretical methods for their characterization. We use two possible ways
of describing a graph: the Adjacency List and the Nickel Index. Both are briefly
explained in this section.
\subsection{Adjacency List}
\begin{figure}[htb]
\centerline{%
\includegraphics[width=0.3\textwidth]{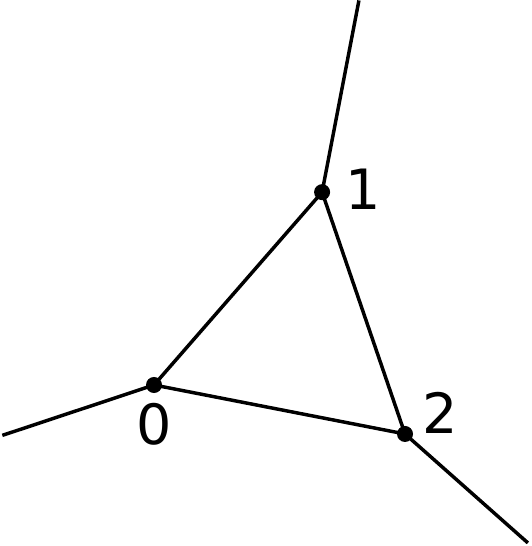}}
\caption{Triangle diagram with the Nickel Index: \( e12|e2|e| \) and an Adjacency List: \( [-1, 0], [0, 1], [0, 2], [-1, 1], [1, 2], [2, -1] \).}
\label{Fig:triangle}
\end{figure}

To get the Adjacency List for a graph we have to number the vertices
with integers. Next, we write down every connection (line) that exists
among the vertices. For example, the graph of Fig.~\ref{Fig:triangle}
would have the following Adjacency List: \( [-1, 0], [0, 1], [0, 2],
[-1, 1], [1, 2], [2, -1] \).  Each of the square brackets represents a
line. For example \( [1,2] \) represents the connection between vertex 1
and vertex 2. The order of the connections is not important. External
fields are denoted with \( -1 \). In the current version of Loopedia,
external fields \emph{cannot} be numbered with \(  -2, -3, -4,  \) \dots,
only with \(-1\).

The Adjacency List depends on the numbering of the vertices so that there is no
obvious way to obtain a canonical (unique) one.
\subsection{Nickel Index}
The Nickel Index consists of several parts --- one part for each vertex.
Each part is closed with a vertical bar \( | \) (pipe symbol). Let us
consider again the diagram from Fig.~\ref{Fig:triangle}. It has the
Nickel Index \( \mathbf{e12|e2|e|} \). To construct this index, we first
have to number the vertices with integers, starting at zero. Here, we
have three vertices so that the labels are 0, 1, 2.

The first part of the Index (read the usual way from left to right)
reads \( e12| \) and lists just all connections that vertex number 0
has. Here, \( e \) stands for an external field. So this part tells us
that vertex 0 is connected to vertex 1, vertex 2, and an external field.

The second part of the index, corresponding to vertex number 1, reads \( e2| \)
and lists all \emph{new} connections of vertex 1 that were not stated in the
previous part. In this example, the connection from 0 to 1 was already mentioned.

The third and last part, describing the connections of vertex 2, reads \( e| \)
and lists the one external connection of this vertex that was not mentioned by
the previous parts.

\begin{figure}[htb]
\centering
\begin{subfigure}[b]{0.49\textwidth}
  % \centering
  \hspace{0.13\linewidth}
  \includegraphics[width=0.5\linewidth]{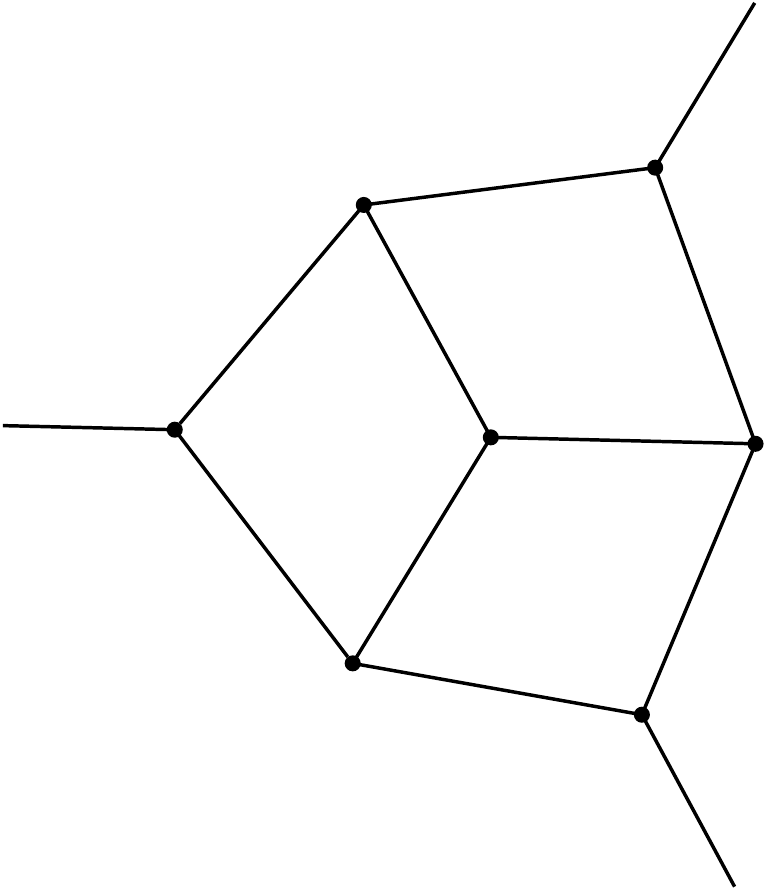}
  \caption{Nickel Index: \( e12|34|35|6|e6|e6|| \)\\
      Adjacency list:\\
      \( [-1,1],[1,2],[1,4],[2, 3],[3,4],[4,6], \)\\
      \( [3,5],[2,7],[6,-1],[6,5],[5,7],[7,-1] \)}
  \label{Fig:examples_1}
\end{subfigure}
\begin{subfigure}[b]{0.49\textwidth}
  % \centering
  \hspace{0.08\linewidth}
  \includegraphics[width=0.7\linewidth]{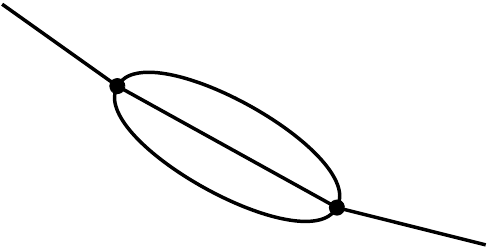}
  \caption{Nickel Index: \( e111|e| \)\\
  Adjacency list:\\
  \( [-1, 1], [1, 2], [1, 2], [1, 2], [2, -1] \)}
  \vspace{1em}
\end{subfigure}
\caption{Further examples}
\label{Fig:examples}
\end{figure}

Note that each line of the graph is represented by exactly one character of the
index. Also, as each vertex has to be represented by one part, the number of
vertical bars in the index is exactly the same as the number of vertices. This
way, it may happen that a particular vertex has no new connections that we can
mention in its corresponding part of the Nickel Index. Thus, we just write a
closing vertical bar. An example of this case is shown in
Fig.~\ref{Fig:examples_1}.

This index depends highly on the way we number the vertices. The crucial
part is that if we write down all possible indices (for all numberings)
according to the presented rules, there will be exactly one of them which is
minimal in a specific sense. This minimal index is called the real
\emph{Nickel~Index} and can be used to extract many properties of the graph.

Two further example graphs and their corresponding Adjacency List and Nickel
Index are shown in Fig.~\ref{Fig:examples}.

A more detailed description of the Nickel Index can be found in the paper of
Batkovich~et~al.~\cite{Batkovich:2014bla}, where they introduce the
\texttt{GraphState} library. It is used by Loopedia to manipulate Nickel
Indices.
\section{First version of Loopedia}
\begin{figure}[tb]
\centering
\fbox{%
\includegraphics[width=0.55\textwidth]{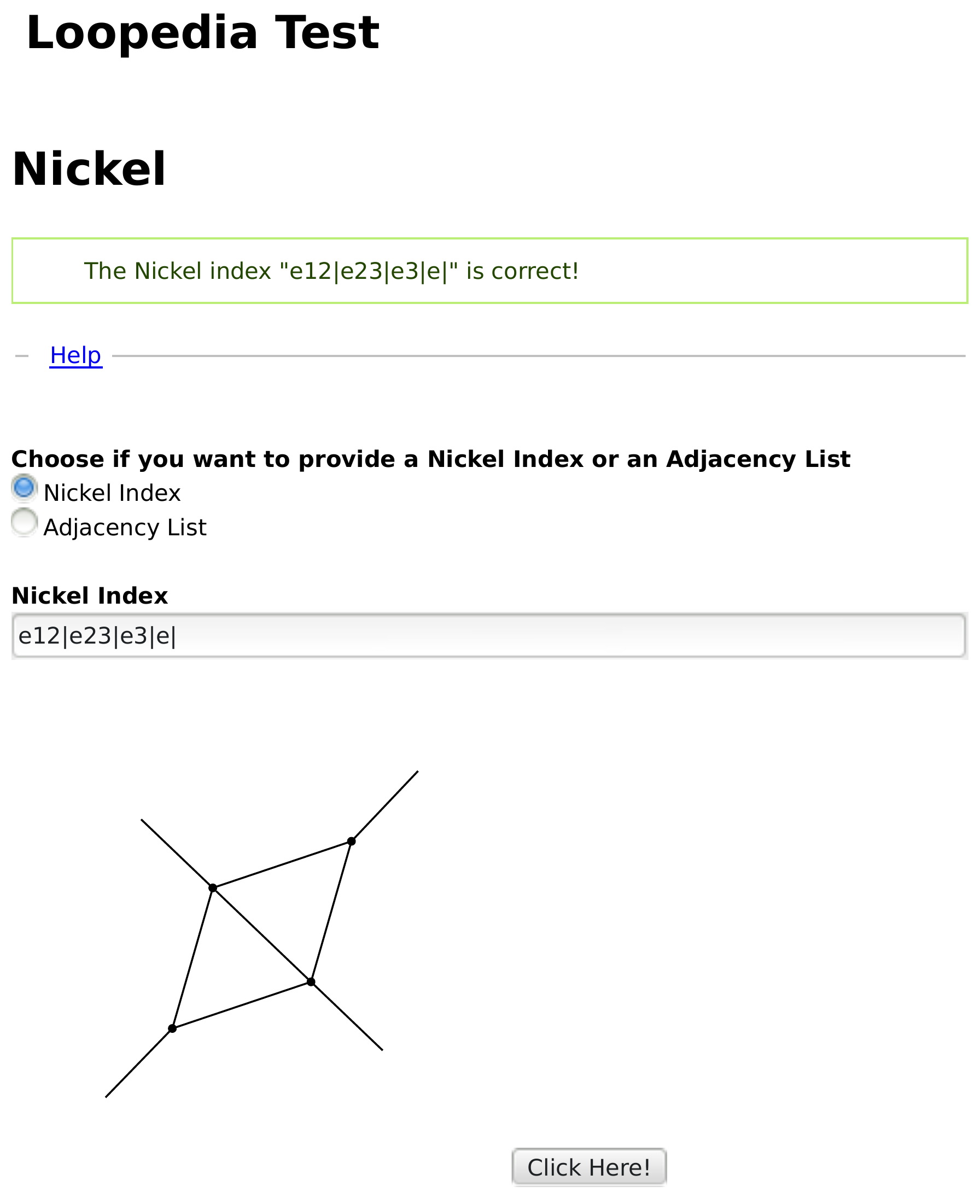}
}
\caption{If you select ``Nickel Index'', type it in and hit Enter-Key.}
\label{Fig:scr01}
\end{figure}

\begin{figure}[tb]
\centering
\fbox{%
\includegraphics[width=0.55\textwidth]{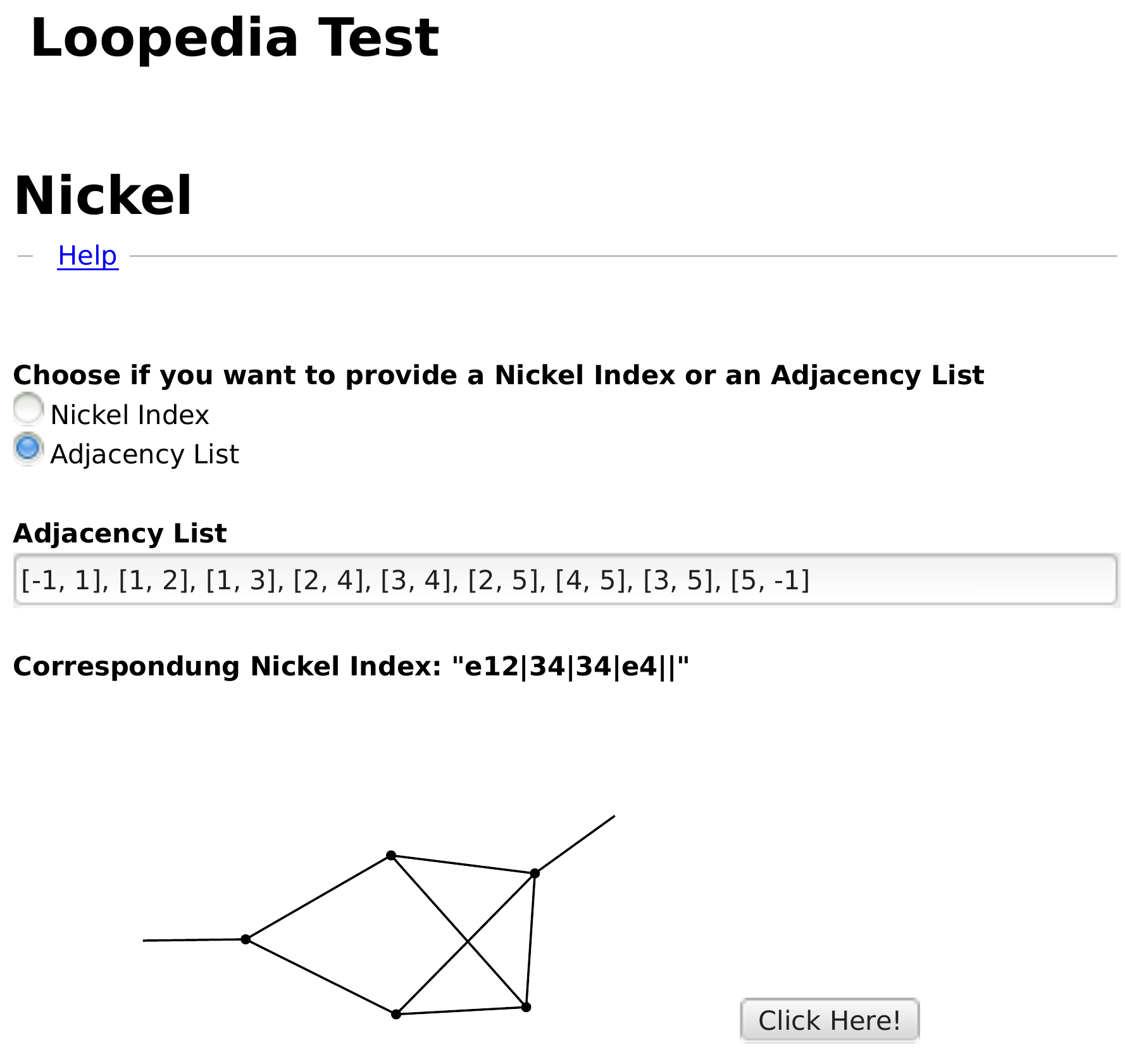}
}
\caption{If you select ``Adjacency List'', type it in and click on ``Click Here!''}
\label{Fig:scr02}
\end{figure}

The first version of Loopedia is already online and can be found at this
address: \url{http://www.mpp.mpg.de/~papara/loopediatest}

At this stage, the user can type in an Adjacency List or a Nickel Index
and the website will generate the corresponding graph using
\texttt{Graphviz}~\cite{SPE:SPE338}. If an Adjacency List was given, it
will also display the corresponding Nickel Index. See screen-shots in
Figs.~\ref{Fig:scr01} and \ref{Fig:scr02}.

\section{Conclusion}
The Loopedia project, a public database of loop integrals, was introduced. In
order to characterize those integrals two notations were presented: the
\mbox{Adjacency} List and the Nickel Index. A first version of the Loopedia
website was demonstrated. It allows for automatic graph generation and
conversion from Adjacency List to Nickel Index.

The next step is to implement an upload function so that the database can be
filled and tested.

\vspace{3em}
I thank the following people for very useful discussions: Christian Bogner,
Sophia Borowka, Thomas Hahn, Gudrun Heinrich, Stephen Jones, Matthias Kerner,
Andreas von Manteuffel, and Erik Panzer.
%

%uncomment the following lines to place a figure
%\begin{figure}[htb]
%\centerline{%
%\includegraphics[width=12.5cm]{Fig1}}
%\caption{Plot of ...}
%\label{Fig:F2H}
%\end{figure}


\begin{thebibliography}{99}

% 1
%\cite{Hahn:1998yk}
\bibitem{Hahn:1998yk}
  T.~Hahn and M.~Perez-Victoria,
  %``Automatized one loop calculations in four-dimensions and D-dimensions,''
  Comput.\ Phys.\ Commun.\  \textbf{118} (1999) 153
  \texttt{[arXiv:hep-ph/9807565]}.
  %%CITATION = HEP-PH/9807565;%%
  %1045 citations counted in INSPIRE as of 05 Oct 2015

% 2
%\cite{Binoth:2008uq}
\bibitem{Binoth:2008uq}
  T.~Binoth, J.-P.~Guillet, G.~Heinrich, E.~Pilon and T.~Reiter,
  %``Golem95: A Numerical program to calculate one-loop tensor integrals with up to six external legs,''
  Comput.\ Phys.\ Commun.\  \textbf{180} (2009) 2317
  \texttt{[arXiv:0810.0992 [hep-ph]]}.
  %%CITATION = ARXIV:0810.0992;%%
  %134 citations counted in INSPIRE as of 06 Oct 2015


% 3
%\cite{ValeryPhD}
\bibitem{ValeryPhD}
  Yundin, Valery: Massive loop corrections for collider physics; Dissertation, Humboldt-Universit\"at zu Berlin, urn:nbn:de:kobv:11-100199626


% 4
%\cite{Fleischer:2010sq}
\bibitem{Fleischer:2010sq}
  J.~Fleischer and T.~Riemann,
  %``A Complete algebraic reduction of one-loop tensor Feynman integrals,''
  Phys.\ Rev.\ D \textbf{83} (2011) 073004
  \texttt{[arXiv:1009.4436 [hep-ph]]}.
  %%CITATION = ARXIV:1009.4436;%%
  %36 citations counted in INSPIRE as of 06 Oct 2015

% 5
%\cite{Ellis:2007qk}
\bibitem{Ellis:2007qk}
  R.~K.~Ellis and G.~Zanderighi,
  %``Scalar one-loop integrals for QCD,''
  JHEP {\bf 0802} (2008) 002
  \texttt{[arXiv:0712.1851 [hep-ph]]}.
  \url{http://qcdloop.fnal.gov/}
  %%CITATION = ARXIV:0712.1851;%%
  %248 citations counted in INSPIRE as of 05 Oct 2015


% 6
%\cite{vanHameren:2010cp}
\bibitem{vanHameren:2010cp}
  A.~van Hameren,
  %``OneLOop: For the evaluation of one-loop scalar functions,''
  Comput.\ Phys.\ Commun.\  \textbf{182} (2011) 2427
  \texttt{[arXiv:1007.4716 [hep-ph]]}.
  %%CITATION = ARXIV:1007.4716;%%
  %127 citations counted in INSPIRE as of 06 Oct 2015

% 7
%\cite{Buckley:2006nm}
\bibitem{Buckley:2006nm}
  A.~Buckley, M.~R.~Whalley, W.~J.~Stirling, J.~M.~Butterworth, E.~Nurse and B.~Waugh,
  %``HepForge: A Lightweight development environment for HEP software,''
  \texttt{arXiv:hep-ph/0605046}.
  \url{https://www.hepforge.org/}
  %%CITATION = HEP-PH/0605046;%%

% 8
%\cite{Batkovich:2014bla}
\bibitem{Batkovich:2014bla}
  D.~Batkovich, Y.~Kirienko, M.~Kompaniets and S.~Novikov,
  %``GraphState - a tool for graph identification and labelling,''
  \texttt{arXiv:1409.8227 [hep-ph]}.
  %%CITATION = ARXIV:1409.8227;%%
  %3 citations counted in INSPIRE as of 24 sept. 2015

% 9
\bibitem{SPE:SPE338}
  Gansner, E. R. and North, S. C. (2000),
  % An open graph visualization system and its applications to software engineering.
  Softw: Pract. Exper., 30: 1203--1233.
  \url{www.graphviz.org}
\end{thebibliography}
\end{document}